\documentclass[conference]{IEEEtran}
\IEEEoverridecommandlockouts
\usepackage{cite}
\usepackage{amsmath,amssymb,amsfonts}
\usepackage{algorithmic}
\usepackage{graphicx}
\usepackage{textcomp}
\usepackage{xcolor}
\usepackage{comment}
\usepackage{multirow}
\usepackage{cite}
\usepackage{subcaption}
\usepackage{booktabs}
\usepackage{array}
\usepackage{tikz}        
\def\BibTeX{{\rm B\kern-.05em{\sc i\kern-.025em b}\kern-.08em
    T\kern-.1667em\lower.7ex\hbox{E}\kern-.125emX}}
\begin{document}

\title{EchoFree: Towards Ultra Lightweight and Efficient Neural Acoustic Echo Cancellation
}




\author{
\IEEEauthorblockN{
Xingchen Li\IEEEauthorrefmark{2},
Boyi Kang\IEEEauthorrefmark{2},
Ziqian Wang,
Zihan Zhang,
Mingshuai Liu,
Zhonghua Fu\IEEEauthorrefmark{1},
Lei Xie
}
\IEEEauthorblockA{
Audio, Speech and Language Processing Group (ASLP@NPU), School of Computer Science \\
Northwestern Polytechnical University, Xi’an, China \\
\{lixingchen, beaukang, zq\_wang, zhzhang, liumingshuai\}@mail.nwpu.edu.cn, \{mailfzh, lxie\}@nwpu.edu.cn
}

\thanks{\IEEEauthorrefmark{2}Equal contribution. \IEEEauthorrefmark{1}Corresponding author.}

}



\maketitle

\begin{abstract}
In recent years, neural networks (NNs) have been widely applied in acoustic echo cancellation (AEC). However, existing approaches struggle to meet real-world low-latency and computational requirements while maintaining performance. To address this challenge, we propose EchoFree, an ultra lightweight neural AEC framework that combines linear filtering with a neural post filter. Specifically, we design a neural post-filter operating on Bark-scale spectral features. Furthermore, we introduce a two-stage optimization strategy utilizing self-supervised learning (SSL) models to improve model performance. We evaluate our method on the blind test set of the ICASSP 2023 AEC Challenge. The results demonstrate that our model, with only \textbf{278K} parameters and \textbf{30 MMACs} computational complexity, outperforms existing low-complexity AEC models and achieves performance comparable to that of state-of-the-art lightweight model DeepVQE-S. The audio examples are available \footnote{https://echofree2025.github.io/EchoFree-demo/}.
\end{abstract}

\begin{IEEEkeywords}
component, formatting, style, styling, insert
\end{IEEEkeywords}

\section{Introduction}
Acoustic echo cancellation (AEC) is a critical front-end task in far-field and hands-free speech communication systems. It aims to suppress echoes resulting from the feedback of far-end speech captured by local microphones, which can severely degrade user experience and downstream performance in automatic speech recognition (ASR) and speaker verification systems. Typically, AEC systems are categorized into traditional digital signal processing (DSP)-based approaches, modern neural network (NN)-based methods, and hybrid systems that combine both.


Traditional digital signal processing (DSP) based AEC approaches employ adaptive filtering algorithms to estimate the near-end speech \cite{DBLP:journals/taslp/Duttweiler00} or echo path \cite{DBLP:journals/ejasp/LuoYKYY24, DBLP:journals/tsp/SooP90, DBLP:journals/taslp/PaleologuBC13}. Such algorithms struggle to handle nonlinear echoes \cite{DBLP:conf/icassp/BenderskySM08} and require residual echo cancellation algorithms, resulting in poor generalization ability in complex real-world scenarios.


Recent advancements in deep learning have significantly
improved AEC performance, enabling the suppression of
nonlinear echo components and achieving superior results
compared to conventional linear filtering techniques\cite{DBLP:journals/corr/abs-2005-09237, DBLP:conf/asru/ZhangYZYW23, DBLP:conf/interspeech/ZhangKLHX21, DBLP:conf/icassp/ZhangWYW22, DBLP:conf/icassp/ZhangWSFTFX22, DBLP:conf/icassp/SunLLLJL23, DBLP:journals/corr/abs-2105-14666, DBLP:conf/interspeech/RisteaISPGC23, DBLP:conf/icassp/ValinTHIK21, DBLP:conf/iwaenc/ShetuDAHM24, DBLP:journals/corr/abs-2410-13620, DBLP:conf/interspeech/ChenY0GLLW23}. Nevertheless, existing deep learning-based AEC models are often characterized by large parameter sizes and high computational complexity, posing significant challenges for deployment on resource-constrained devices. 

\begin{figure}[ht]
  \centering
  \includegraphics[width=0.45\textwidth]{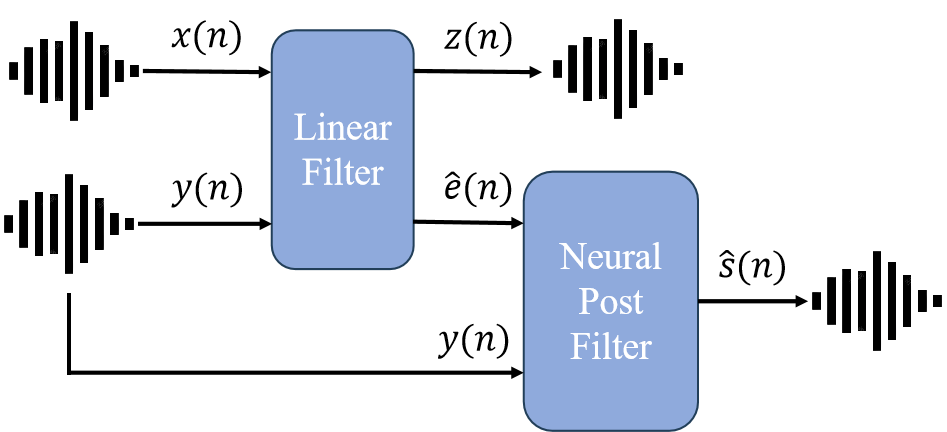}  
  \caption{The hybrid approach combines a linear acoustic echo canceller with a proposed neural post filter designed for residual echo suppression. Model details are provided in \figurename~\ref{fig:three_subfigs}. }
  \label{fig:pipeline}
\end{figure}



To address these challenges, we propose EchoFree, an ultra lightweight AEC model designed for real-world applications on resource-limited edge devices. EchoFree adopts a hybrid and efficient approach that combine linear filtering with a neural post filter shown in \figurename~\ref{fig:pipeline}. The neural post filter leverages a compressed power spectrum representation in the Bark-scale, which has been proven to effectively reduce computational complexity while preserving critical speech information\cite{DBLP:conf/icassp/ValinTHIK21}. Furthermore, the embeddings produced by self-supervised learning (SSL) models have been proved to have rich acoustic and semantic information\cite{vectok}, which has been adopted to further improve the capabilities of speech enhancement models\cite{DBLP:conf/icassp/Shankar0X024}. Inspired by this, we propose and apply a two-stage optimization strategy based on a SSL model. By utilizing the SSL model to guide the neural network in a progressive learning manner—from coarse-grained to fine-grained spectral representations—the proposed method enhances the model's ability to perform effective echo suppression.

In this work, we perform comprehensive ablation studies to validate the effectiveness of the proposed two-stage optimization strategy, and systematically compare EchoFree with several representative state-of-the-art AEC models in terms of performance and efficiency. Experimental results demonstrate that EchoFree achieves performance comparable to the state-of-the-art DeepVQE-S, using only \textbf{278K} parameters and \textbf{30 MMACs/s}, establishing a superior trade-off between efficiency and performance.

\section{Related Work}
\subsection{DSP-based method}
Traditional DSP based AEC approaches employ adaptive filtering algorithms to estimate the near-end speech \cite{DBLP:journals/taslp/Duttweiler00} or echo path \cite{DBLP:journals/ejasp/LuoYKYY24, DBLP:journals/tsp/SooP90, DBLP:journals/taslp/PaleologuBC13}, such as the normalized least mean squares (NLMS), affine projection (APA), or recursive least squares (RLS) algorithms. However, these algorithms inherently assume a linear and stationary echo path, making them inadequate for handling nonlinear distortions introduced by loudspeakers or hardware imperfections. These methods typically rely on handcrafted residual echo suppression modules to mitigate remaining echo components \cite{DBLP:conf/icassp/BenderskySM08}, but such modules struggle to distinguish residual echo from speech, often leading to speech distortion or echo leakage. The problem is inherently challenging due to the dynamic nature of acoustic environments, including time-varying echo paths caused by changes in room geometry, user movement, or device placement. As a result, DSP-based AEC systems exhibit poor generalization in real-world scenarios with dynamic acoustics and varying noise conditions. 

\subsection{NN-based method}
On the other hand, recent advancements in deep learning have significantly improved AEC performance, enabling the suppression of nonlinear echo components and achieving superior results compared to conventional linear filtering techniques \cite{DBLP:journals/corr/abs-2005-09237, DBLP:conf/asru/ZhangYZYW23, DBLP:conf/interspeech/ZhangKLHX21, DBLP:conf/icassp/ZhangWYW22, DBLP:conf/icassp/ZhangWSFTFX22, DBLP:conf/icassp/SunLLLJL23, DBLP:journals/corr/abs-2105-14666, DBLP:conf/interspeech/RisteaISPGC23, DBLP:conf/icassp/ValinTHIK21, DBLP:conf/iwaenc/ShetuDAHM24, DBLP:journals/corr/abs-2410-13620, DBLP:conf/interspeech/ChenY0GLLW23}. 
A common strategy involves combining DSP methods with neural networks to form hybrid frameworks. 
For example, Ma et al. \cite{DBLP:journals/corr/abs-2005-09237} proposed a system where an adaptive digital filter removes most of the linear echo, while a recurrent neural network (RNN) targets the residual nonlinear components.
To enhance temporal modeling capabilities, Zhang et al. \cite{DBLP:conf/interspeech/ZhangKLHX21} introduced a complex-valued neural network architecture that employs frequency-time LSTM (F-T-LSTM) layers, effectively capturing dependencies across both time and frequency domains. 

In addition to hybrid frameworks, end-to-end neural architectures have also demonstrated promising performance in acoustic echo cancellation. 
Zhang et al. \cite{DBLP:conf/icassp/ZhangWSFTFX22} introduced a multi-task residual echo suppression framework that cascades a linear AEC front-end with a gated convolutional F-T-LSTM neural post-filter, trained jointly with a voice activity detection (VAD) module and optimized using an echo-aware loss function to enhance echo suppression while mitigating speech distortion. 
Ma et al. \cite{DBLP:journals/corr/abs-2105-14666} proposed EchoFilter, an end-to-end neural AEC system that directly operates on time-domain waveforms using temporal convolution and LSTM modules, and incorporates a local attention mechanism to address time delays and reverberation, with multitask learning employed to enhance robustness under double-talk and nonlinear distortion. 
Indenbom et al. \cite{DBLP:conf/interspeech/RisteaISPGC23} proposed DeepVQE, a unified real-time speech enhancement framework that integrates acoustic echo cancellation, noise suppression, and dereverberation using a residual convolutional neural network (CNN) combined with recurrent layers and a cross-attention mechanism to jointly model multiple interference sources in the time-frequency domain. 

\begin{figure*}[ht]
  \centering
  \begin{subfigure}{\textwidth}
      \centering
      \includegraphics[width=0.9\textwidth]{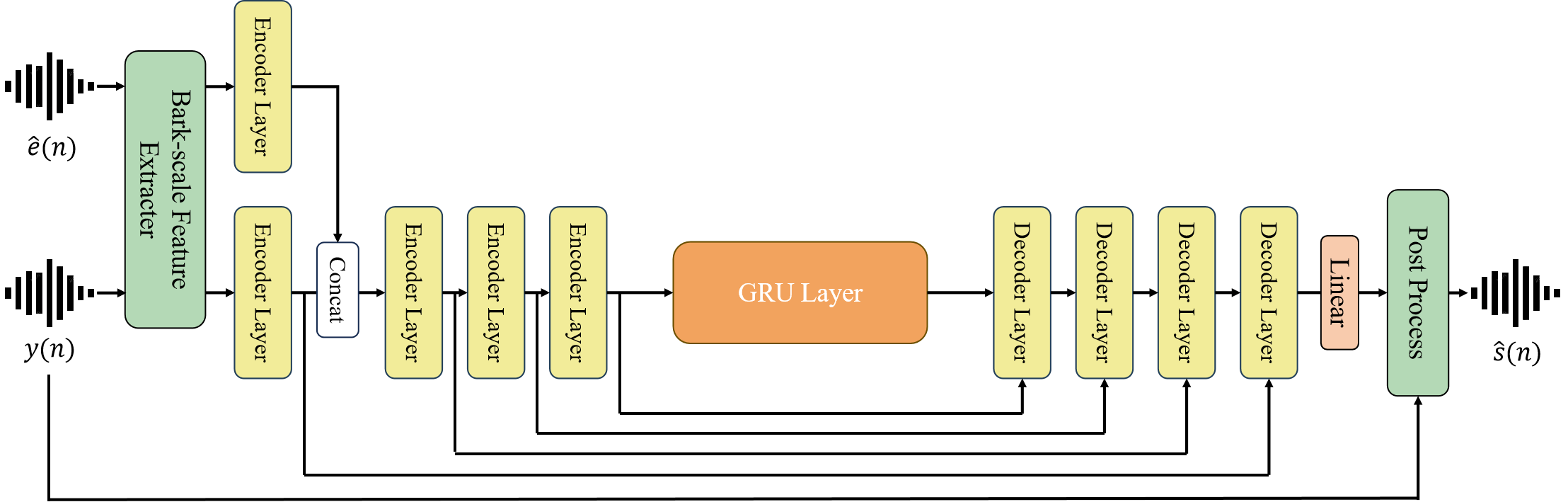}  
      \caption{Overall architecture of neural post filter.}
      \label{fig:arch_full}
  \end{subfigure}

  \begin{tikzpicture}
    \draw[dashed] (0,0) -- (\linewidth,0);
  \end{tikzpicture}
  
  \begin{subfigure}{0.45\textwidth}
      \centering
      \includegraphics[width=\textwidth]{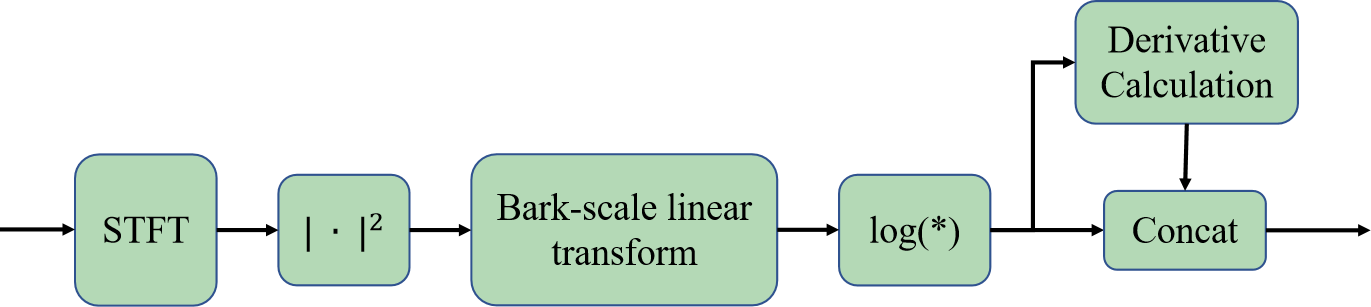}
      \caption{Bark feature extraction module.}
      \label{fig:bark}
  \end{subfigure}
  \begin{tikzpicture}
      \draw[dashed] (0,0) -- (0,3.6);  
  \end{tikzpicture}
  \begin{subfigure}{0.45\textwidth}
      \centering
      \includegraphics[width=\textwidth]{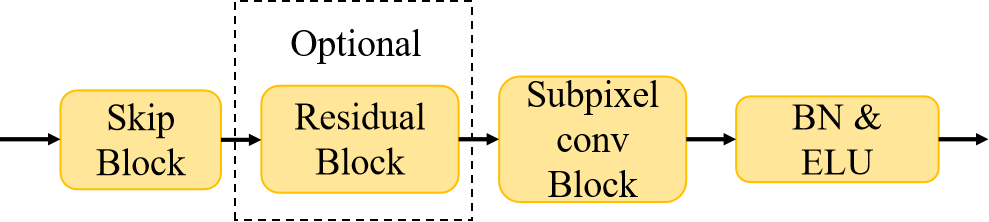}
      \caption{Decoder layer.}
      \label{fig:decoder}
  \end{subfigure}
  
  \caption{Neural network architecture of our proposed neurual post filter. (a) provides an overall architecture of the neural post filter of EchoFree, (b) shows the structure of bark-scale feature extractor, (c) shows the structure of decoder layer.}
  \label{fig:three_subfigs}
\end{figure*}
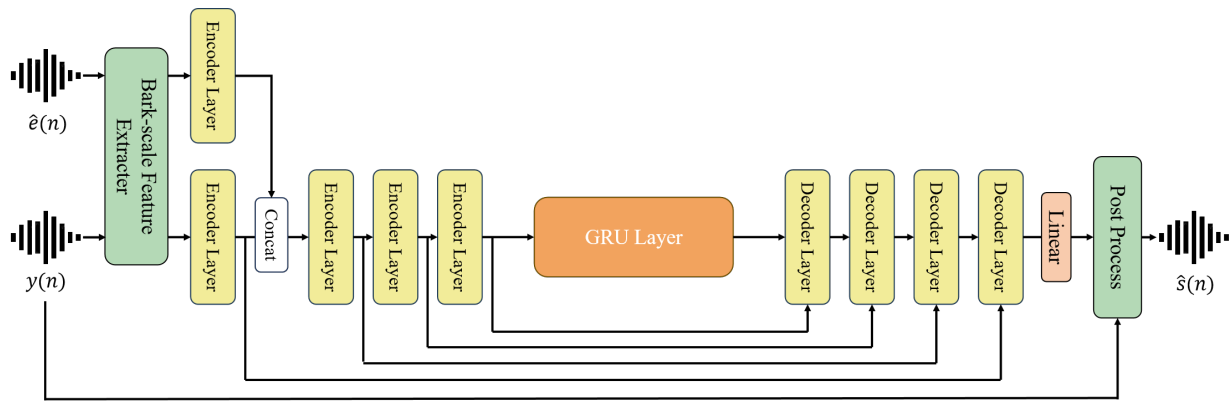
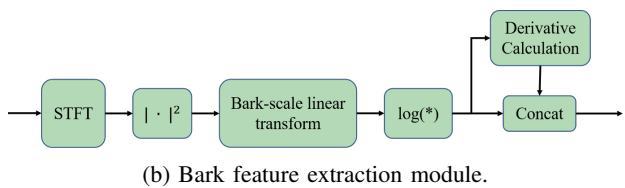
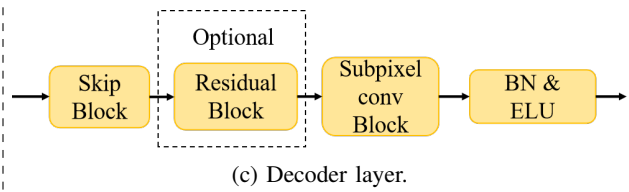

However, existing AEC systems often face a significant trade-off between performance and computational complexity, prompting extensive research into lightweight solutions.
For example, Valin et al. \cite{DBLP:conf/icassp/ValinTHIK21} presented a hybrid architecture that integrates a traditional linear acoustic echo canceller with a lightweight neural post-filter based on PercepNet, designed to jointly suppress residual echo and background noise in real time using perceptually motivated features and low-complexity recurrent-convolutional layers. 
Shetru et al. \cite{DBLP:conf/iwaenc/ShetuDAHM24} proposed a hybrid joint AEC and noise reduction framework that integrates the ultra-low complexity ULCNet model into a traditional linear AEC pipeline, enabling residual echo and noise suppression with minimal computational cost while maintaining competitive performance in low-resource environments. 
Chen et al. \cite{DBLP:conf/interspeech/ChenY0GLLW23} Chen et al. employed a Bark-scale auditory filterbank to enhance the fidelity of the near-end speech, thereby enabling neural post-filtering with extremely low computational complexity.

\section{Proposed Method}

\subsection{Problem Formulation}

We consider a full-duplex communication scenario, where the far-end signal 
$x$ is played through a loudspeaker, undergoes transformations such as room reflections, and is then re-captured by the near-end microphone. The recorded signal consists of two components: near-end signal and far-end echo. Thus, the signal model can be expressed as:
\begin{equation}
    y(n) = s(n) + e(n)
\end{equation}
where $n$ represents the time index of the sample points, $s(n)$ represents the near-end received signal (composed of a mixture of speech and background noise), and $e(n)$ denotes the echo signal captured by the microphone after transformation.

In this work, we focus solely on eliminating the echo signal from the far-end while ignoring background noise. The objective is to estimate the near-end signal $s(n)$ given the mixed microphone signal $y(n)$ and the far-end reference signal $x(n)$. As illustrated in \figurename~\ref{fig:pipeline}, our approach employs a cascaded framework consisting of linear filtering followed by neural post filter. Initially, the linear filter estimates the echo signal $\hat{e}(n)$ by modeling the transformation of the far-end signal $x(n)$, thereby producing the error signal $z(n)$ as follows:
\begin{equation}
\begin{aligned}
\begin{split}
    z(n) &= y(n) - \hat{e}(n)                \\
         &= s(n) + \{e(n) - \hat{e}(n)\}       \\
         &= s(n) + r(n)
\end{split}
\end{aligned}
\end{equation}
where $r(n)$ represents the residual echo, i.e., the discrepancy between the actual echo signal $e(n)$ and the estimated echo signal $\hat{e}(n)$.

The neural post filter module is designed to further refine the echo suppression by estimating the residual echo $r(n)$ based on the mixed microphone signal $y(n)$ and the estimated echo $\hat{e}(n)$ obtained from the linear filtering stage.

\subsection{Linear Filtering}

We use the partitioned-block-frequency-domain-adaptive Kalman filter \cite{DBLP:conf/icassp/KuechME14} as the linear filtering preprocessing module. This linear filtering algorithm takes the mixed microphone signal $y(n)$ and the far-end reference signal $x(n)$ as inputs, producing the estimated echo signal $\hat{e}(n)$ and the residual signal $z(n)$.

\subsection{Neural Post Filter}

\figurename~\ref{fig:arch_full} shows our neural post filter architecture,which consists of a Bark-based feature extractor module, the main structure of the neural network, and a post-processing module.

\noindent\textbf{Feature Extraction Module:} We first apply the short-time Fourier transform (STFT) to the mixed microphone signal $y(n)$ and the estimated echo $\hat{e}(n)$, obtaining their corresponding spectrogram matrices $Y$ and $\hat{E}$ with dimensions [T, F]. Following the Bark-scale partitioning method from \cite{DBLP:journals/corr/abs-2005-09237}, we compute the logarithmic Bark-scale power spectrum of these spectrograms, compressing them into lower-dimensional feature maps. As shown in \figurename~\ref{fig:bark}, the squared magnitude spectrum of the input audio is computed initially. This spectrum is then multiplied by the mapping matrix $ B $. Finally, the logarithm of the result from the linear transformation is computed, producing the logarithmic power spectrum on the Bark scale. Additionally, as in \cite{DBLP:journals/corr/abs-2005-09237}, we incorporate the first- and second-order derivatives of the Bark-scale feature maps to provide the model with sufficient prior knowledge while maintaining a compact architecture. The Bark-scale power spectrum features and their derivatives are concatenated along the feature dimension, forming the final input representation of shape [T, D].

\noindent\textbf{Neural Network Architecture:} \figurename~\ref{fig:arch_full} shows our neural network architecture, which improves upon previous work using fully connected layers and stacked GRU layers \cite{DBLP:journals/corr/abs-2005-09237, DBLP:conf/icassp/SeidelMF24}. We adopt a new U-Net \cite{DBLP:conf/miccai/RonnebergerFB15} structure as the core of the neural network. This architecture has been proven to achieve better performance with a smaller computational cost. The neural network consists of an encoder/bottleneck layer and a decoder.

The encoder includes two branches. The branch for the mixed microphone signal consists of 4 depthwise separable convolution layers \cite{DBLP:conf/miccai/RonnebergerFB15} with filter sizes of 8, 16, 24, 32. The branch for the echo signal consists of a single depthwise separable convolution layer with 8 filters. As shown in \ref{fig:arch_full}, after the echo signal is encoded, its features are concatenated with the features from the mixed microphone signal branch and passed through the subsequent encoder layers.

To balance performance and computational efficiency, the bottleneck layer includes a unidirectional GRU instead of a more complex LSTM module followed by a linear layer. The decoder is composed of four decoder modules, as shown in \figurename~\ref{fig:decoder}. Similar to \cite{DBLP:conf/interspeech/RisteaISPGC23}, we employ a skip-block mechanism for skip connections. Additionally, we incorporate an optional Residual Block to enhance upsampling performance, which is applied only in the last decoder module. For upsampling, we use a SubPixelConv module, which provides lower computational complexity. Finally, BatchNorm \cite{DBLP:conf/icml/IoffeS15} and the Exponential Linear Unit (ELU) activation function \cite{DBLP:journals/corr/ClevertUH15} are applied to the output of each layer. The decoder module filter sizes, mirroring the encoder, are 24, 16, 8, 1.

To predict the frequency band gain in the Bark-scale, we apply a final linear layer followed by a sigmoid activation function, ensuring that the output gain values remain constrained between 0 and 1.

\noindent\textbf{Post Process.}  In the post-processing stage, the gain obtained in the previous phase is multiplied by the transpose of the mapping matrix $ B $, resulting in the magnitude spectrum mask. This mask is then multiplied by the magnitude spectrum of the mixed signal to obtain the estimated magnitude spectrum of the near-end speech, which is used to derive the estimated near-end speech.

\subsection{Loss Function}

\begin{figure}[ht]
  \centering

  \includegraphics[width=0.45\textwidth]{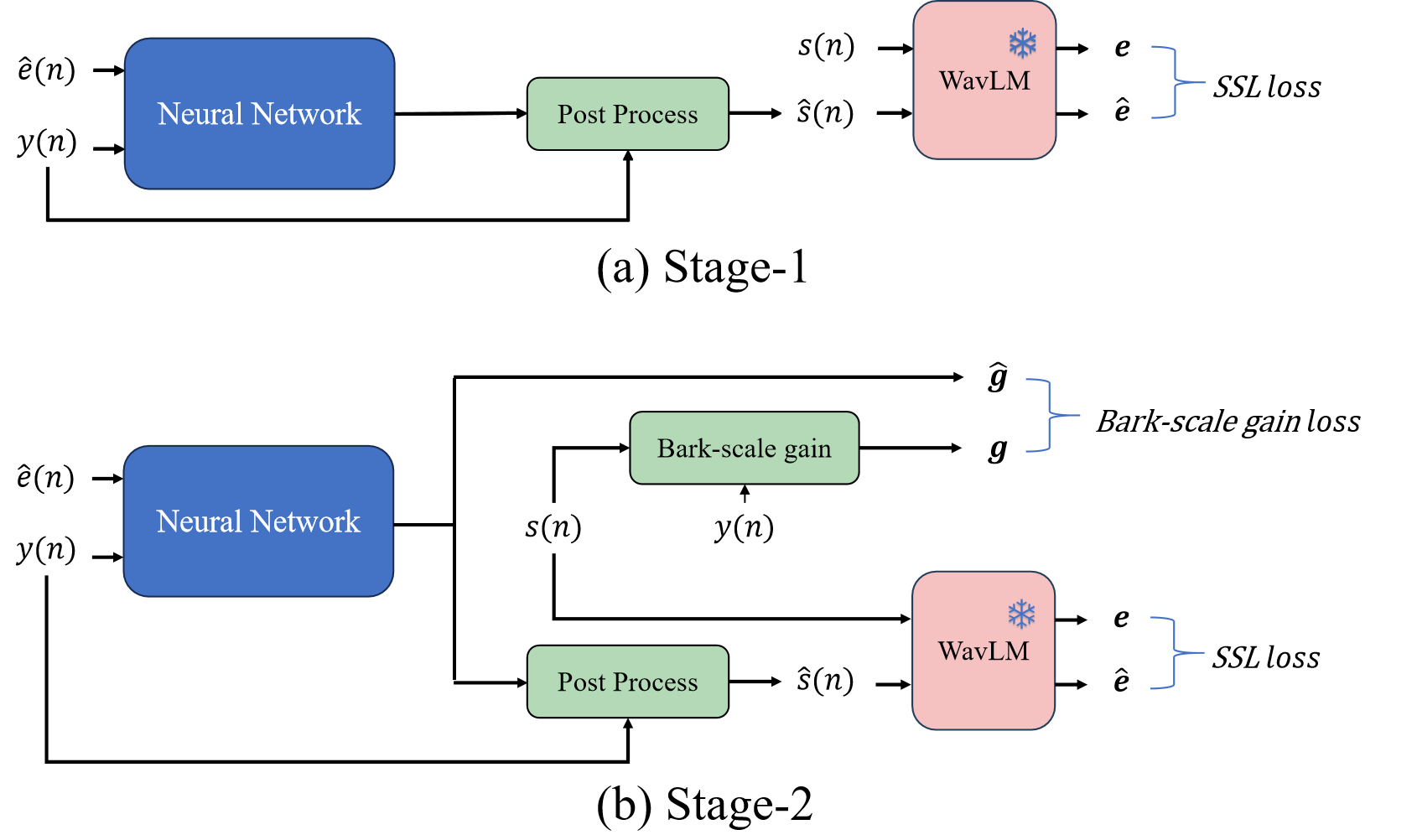}
  \caption{The proposed two-stage training strategy: the first stage uses only the SSL loss, while the second stage employs a combination of Bark-scale gain loss and SSL loss. During training, the parameters of WavLM are kept frozen.}
  \label{fig:training}
  
\end{figure}

The embeddings produced by SSL model have been proved to have rich acoustic and semantic information\cite{vectok}, and various SSL-based training strategies are explored in speech enhancement models\cite{DBLP:conf/icassp/Shankar0X024}. Inspired by this, We adopt a two-stage training strategy, as illustrated in \figurename~\ref{fig:training}, to optimize the model more effectively.
In the first stage, we use an enhanced loss function based on SSL model's embeddings. We refer to this loss as the SSL loss throughout the rest of the paper. Specifically, we compute the MSE loss between the SSL embeddings of the ground truth and the estimated signals as shown in \figurename~\ref{fig:arch_full}. These embeddings are extracted using a pre-trained WavLM-Large model \footnote{https://huggingface.co/microsoft/wavlm-large} \cite{DBLP:journals/jstsp/ChenWCWLCLKYXWZ22}, whose parameters are frozen during training, ensuring that our model aligns effectively with ground truth representations across multiple speech dimensions. The SSL loss is formulated as:
\begin{equation}
    \mathcal{L}_{SSL} = \frac{1}{L} \sum_{l=1}^{L} \|\mathbf{e}_{l} - \mathbf{\hat{e}}_{l}\|^2
\end{equation}
where \(L\) denotes the total number of layers in the WavLM model, \(\hat{e}_{i,l}\) represents the estimated signal embedding for the \(i\)-th sample at the \(l\)-th layer, and \(e_{i,l}\) is the ground truth embedding for the same sample at the \(l\)-th layer.

In the second stage, to encourage fine-grained improvements that are aligned with human auditory perception, we introduce the Bark-scale gain loss $ \mathcal{L}_{Bark} $. This loss function penalizes both the squared and quartic root-mean discrepancies between the predicted gain $ \mathbf{\hat{g}} $ and the target gain $ \mathbf{g} $. Specifically, we formulate the loss as follows:
\begin{equation}
\begin{aligned}
    \mathcal{L}_{Bark} =& 10 (|\mathbf{\hat{g}}|^c - |\mathbf{g}|^c)^4 + (|\mathbf{\hat{g}}|^c - |\mathbf{g}|^c)^2 +  \\
    &0.01 * \text{CrossEntropy}(\mathbf{\hat{g}}, \mathbf{g})
\end{aligned}
\end{equation}
Where c represents the compression coefficient, here $ c=0.5 $. The fourth-order term emphasizes larger deviations in the Bark-scale gain, ensuring significant perceptual errors are strongly penalized. The second-order term provides general stability, while the cross-entropy term adds a regularization effect based on distributional consistency between predicted and target gain patterns.

The second-stage objective is formulated as a weighted combination of the SSL loss and the Bark-scale gain loss. In this stage, the model is optimized toward improving the Bark-scale gain performance, with the SSL loss serving as a regularization term to preserve the learned representation fidelity.

We denote the loss functions used in the two training stages as follows:
\begin{align*}
    \mathcal{L}_{stage\text{-}1} &= \mathcal{L}_{SSL}       \tag{5}\\
    \mathcal{L}_{stage\text{-}2} &= 10 * \mathcal{L}_{Bark} + 0.5 * \mathcal{L}_{SSL} \tag{6}
\end{align*}

\begin{table*}
\caption{Trained on the same dataset, a comparison between EchoFree and state-of-the-art low-complexity AEC models is presented, along with the performance of EchoFree under different training strategies.}
\centering
\begin{tabular}{m{3.5cm} >{\centering\arraybackslash}m{1.2cm} >{\centering\arraybackslash}m{1.2cm} >{\centering\arraybackslash}m{1.2cm} >{\centering\arraybackslash}m{1.2cm} >{\centering\arraybackslash}m{1.2cm} >{\centering\arraybackslash}m{1.2cm}}
\toprule
Method&\# Param.&Macs/s&\shortstack{ST FE \\ EchoMOS}&\shortstack{ST NE \\ DegMOS}&\shortstack{DT \\ EchoMOS}&\shortstack{DT \\ DegMOS} \\
\midrule
ULCNet-AER\cite{DBLP:conf/iwaenc/ShetuDAHM24}&1.12M&173M&2.89&3.04&2.68&\textbf{3.77} \\
Bark-AEC\cite{DBLP:conf/icassp/SeidelMF24}&1.62M&107M&3.16&2.83&2.96&3.27 \\
DeepVQE-S\cite{DBLP:conf/interspeech/RisteaISPGC23}&0.82M&315M&4.13&3.24&\textbf{3.96}&3.69 \\
\midrule
Linear AEC only&-&-&2.91&3.02&2.68&3.76 \\
EchoFree-cost loss only&\textbf{0.28M}&\textbf{30M}&4.15&3.13&3.74&3.52 \\
EchoFree-SSL loss only&\textbf{0.28M}&\textbf{30M}&4.15&3.18&3.91&3.46 \\
EchoFree-proposed&\textbf{0.28M}&\textbf{30M}&\textbf{4.20}&\textbf{3.27}&3.88&3.53 \\
\bottomrule
\label{table:exp}
\end{tabular}
\end{table*}

\section{Experiments and Results}

\subsection{Training Datasets}

We use the clean speech data provided by the ICASSP 2021 DNS Challenge \cite{DBLP:conf/icassp/LiLLZ021} to train the model. We selected nearly 90,000 high-quality clean speech samples with a total duration of 573 hours from this dataset. We used 80,000 samples for training, totaling 506 hours of speech, and the remaining 10,000 samples for the validation set, with a total duration of 67 hours. All speech samples have a sampling rate of 16 kHz.

To enhance acoustic diversity, we adopt a dynamic simulation pipeline. Each training instance is assigned parameters such as signal-to-echo ratio (SER), echo delay, room impulse response (RIR) characteristics, and nonlinear distortion profiles. The signal processing chain operates as follows: 1) Near-end speech: Convolved with randomly selected RIR from acoustic database; 2) Far-end reference: Subjected to nonlinear distortion simulation followed by RIR convolution and time-domain delay ranging from 10ms to 512ms; 3) Signal mixing: Combined under controlled SER ranging from -15 dB to 15 dB.

To account for different communication scenarios, including near-end single talk, far-end single talk, and double talk, we introduce additional randomness into the data generation process. Specifically, the near-end speech signal is set to zero with a 10\% probability to simulate far-end single talk. Similarly, near-end single talk scenarios naturally occur within the duplex samples, eliminating the need for separate explicit inclusion. This dynamic and stochastic simulation strategy ensures that the training data effectively covers a broad range of real-world AEC conditions, thereby enhancing model robustness.

\subsection{Experimental Setup}

For the linear filtering algorithm, we employ the partitioned-block frequency-domain adaptive Kalman filter described in \cite{DBLP:conf/icassp/KuechME14}. Throughout both training and inference, we consistently use 10 partitions and set the FFT length to 256 samples for models utilizing linear filtering as a preprocessing step.

In the feature extraction module, for 16 kHz audio signals, we apply the STFT transformation with a window length of 512 samples, a hop size of 256 samples, and an FFT length of 512 samples, yielding an STFT spectrum with 257 frequency bins. To compress the input feature dimension, we extract Bark-scale features using 100 Bark-scale filters. Additionally, following \cite{DBLP:journals/corr/abs-2005-09237}, we incorporate the first- and second-order derivatives of the first six features. Consequently, the total input feature dimension is 112.

In the neural network module, the microphone branch encoder consists of four encoder layers, each with a convolution kernel size of (4, 3) and a stride of (4, 3). The reference branch encoder includes a single convolutional layer with the same kernel size and stride. The bottleneck layer comprises a GRU with 192 units and a fully connected layer with 192 units. The decoder consists of four decoder modules, where both the skip block and sub-pixel block utilize $1\times1$ convolution kernels. The final decoder module also includes a residual convolution module.

For EchoFree training, we adopt the Adam optimizer \cite{DBLP:journals/corr/KingmaB14} with an initial learning rate of 0.001, which is reduced by a factor of 0.5 if the validation loss does not improve for five consecutive epochs and the minimum value of the learning rate is set to $10^{-5}$. We use a batch size of 128 and train on 10-second speech segments. Training continues until the validation loss fails to improve for ten consecutive epochs. In the second stage, we resume training with a reduced learning rate while keeping the same optimizer, stopping when the validation loss again stagnates for ten epochs. For the SSL model, the pre-trained WavLM-large is employed to compute the SSL loss.

\begin{figure*}
    \centering
    \begin{subfigure}[b]{0.25\textwidth}
        \centering
        \includegraphics[width=\textwidth]{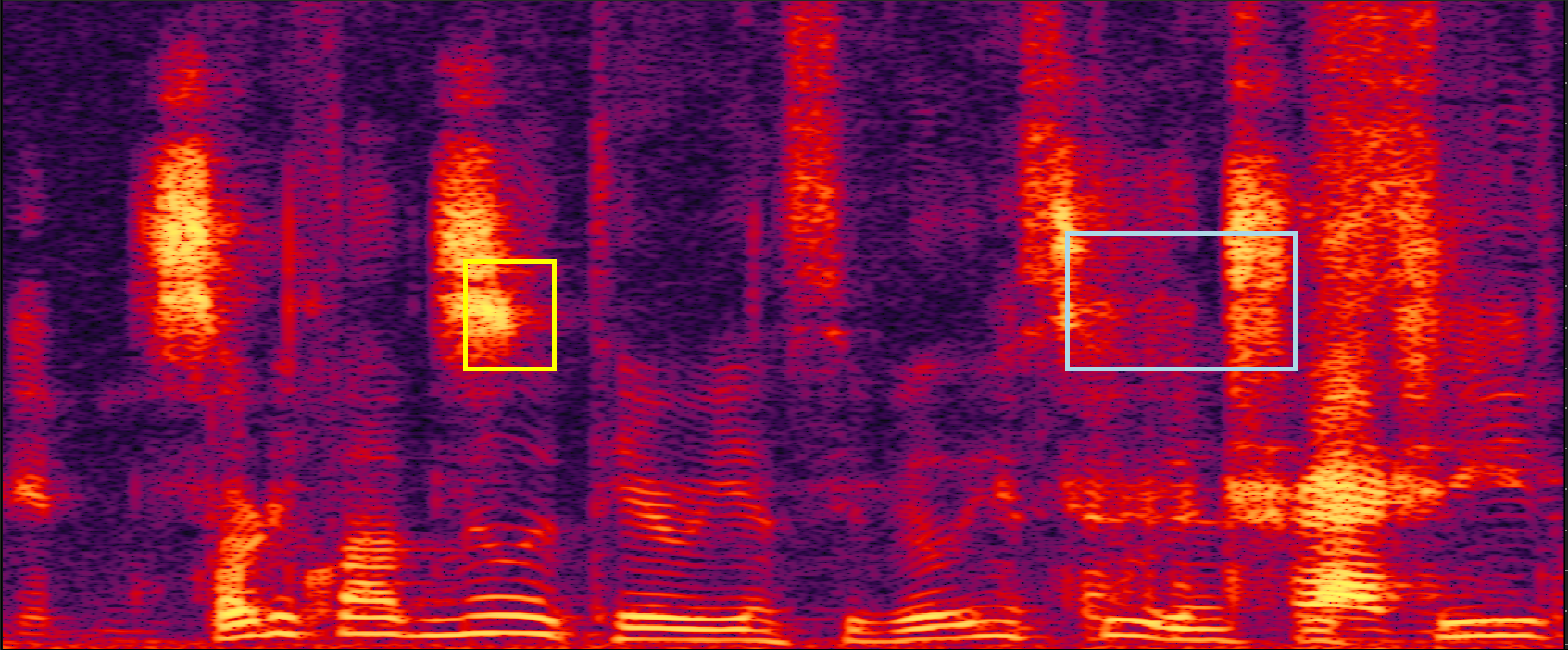}
        \caption{input}
        \label{fig:sub1}
    \end{subfigure}
    \hfill
    \begin{subfigure}[b]{0.25\textwidth}
        \centering
        \includegraphics[width=\textwidth]{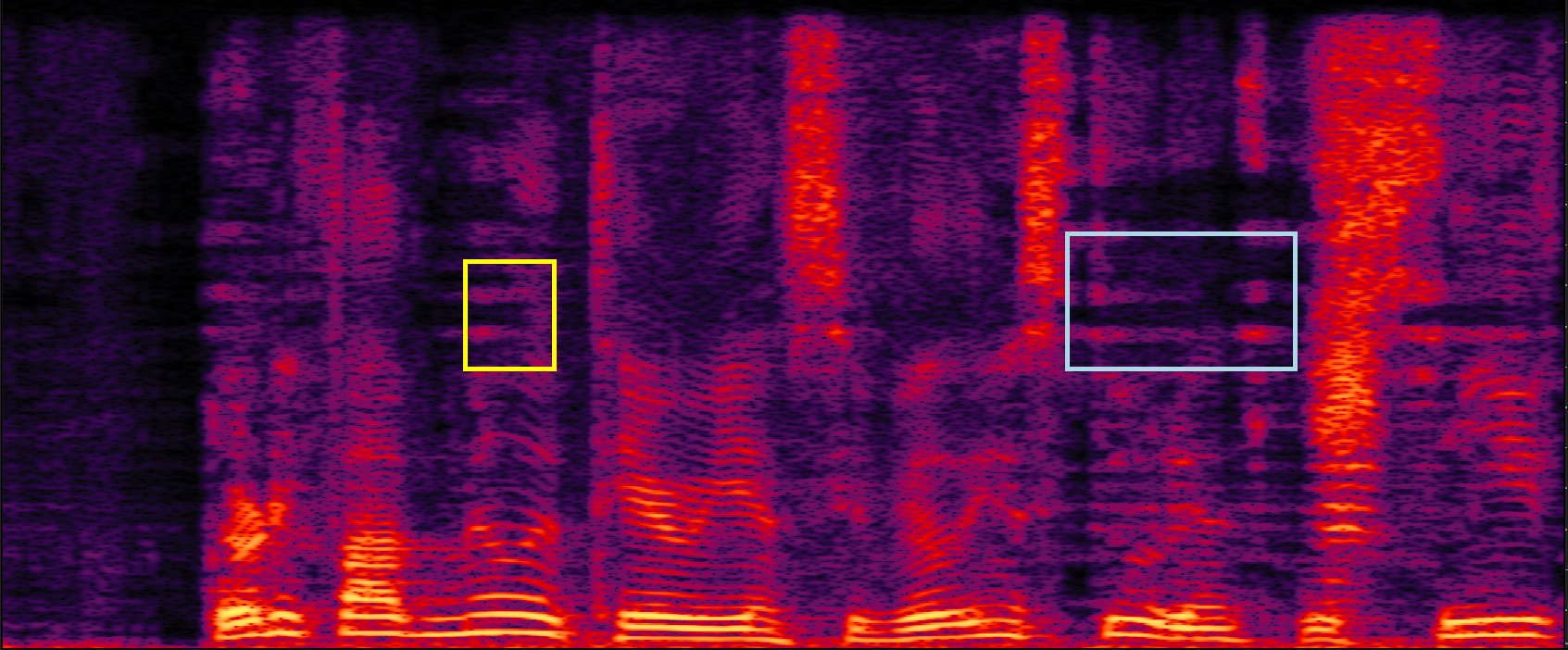}
        \caption{EchoFree stage1}
        \label{fig:sub2}
    \end{subfigure}
    \hfill
    \begin{subfigure}[b]{0.25\textwidth}
        \centering
        \includegraphics[width=\textwidth]{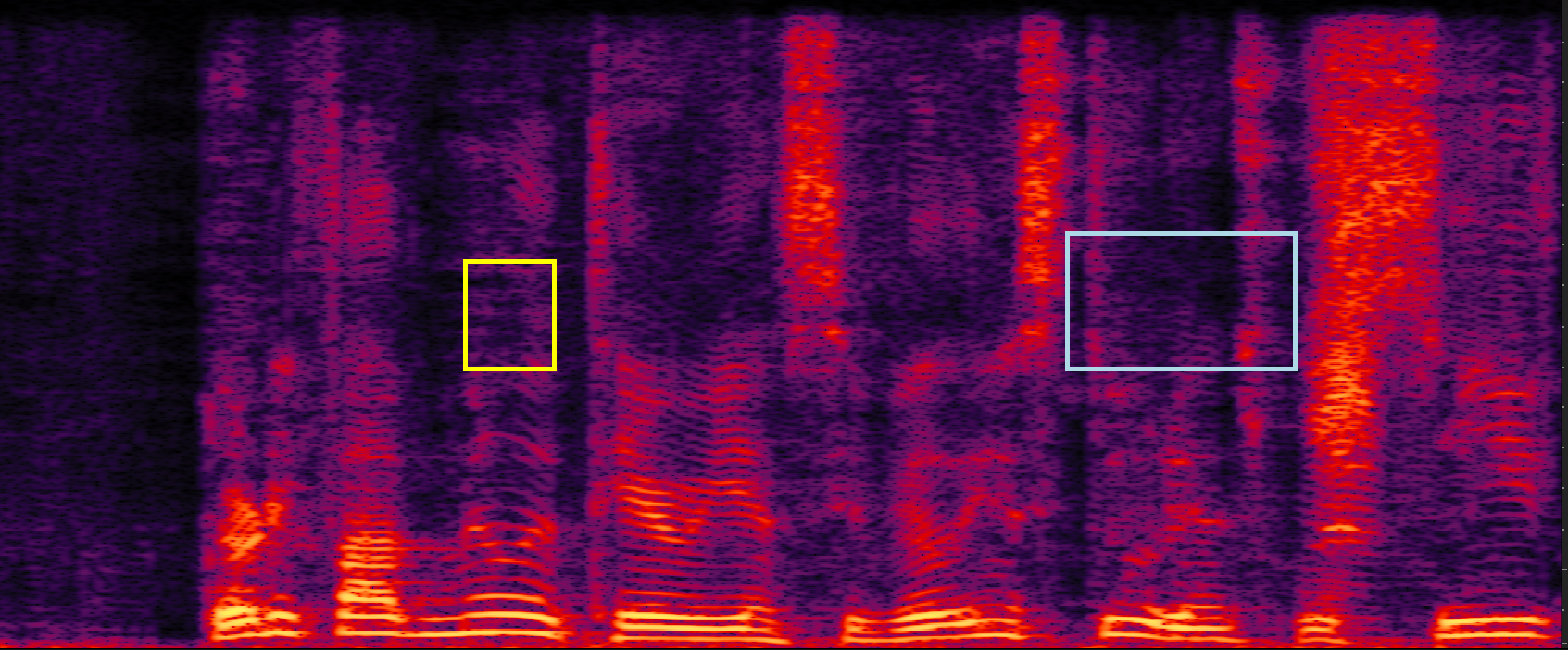}
        \caption{EchoFree stage2}
        \label{fig:sub3}
    \end{subfigure}

    \caption{Visualization of the output at each stage. The input audio (a) contains acoustic echo. The output of the first stage (b) initially suppresses the echo but introduces spectral distortions, which are further corrected in the second stage (c).}
    \label{fig:visualization}
\end{figure*}

\subsection{Evaluation Dataset and Metrics}

We evaluate our method using the blind test set from the ICASSP 2023 AEC Challenge \cite{DBLP:journals/corr/abs-2309-12553}, which comprises 800 test samples. These include 300 double-talk cases, 300 far-end single-talk cases, and 200 near-end single-talk cases. Since our model operates at a 16 kHz sampling rate, we resample the blind test set audio to 16 kHz before processing.

For performance evaluation, we adopt AECMOS \cite{DBLP:conf/icassp/PurinSSSC22}, which includes two key metrics: EchoMOS, assessing echo cancellation performance, and DegMOS, measuring the preservation of near-end speech quality. We follow the open-source AECMOS implementation, using the model \textit{Run\_1668423760\_Stage\_0.onnx} for evaluation. As this model is designed for 48 kHz audio, we upsample our processed outputs to 48 kHz before computing AECMOS scores.

\subsection{Results and Analysis}

To benchmark our approach, we compare it against three state-of-the-art low-complexity AEC models:  
1) ULCNet-AER \cite{DBLP:conf/iwaenc/ShetuDAHM24}, a recently proposed lightweight AEC model optimized for low computational complexity.  
2) Bark-scale feature-based AEC \cite{DBLP:conf/icassp/SeidelMF24}, which employs Bark-scale features as input for the NN post-filter, making it a suitable baseline for evaluating the advantages of our approach. In the following, we refer to this method as Bark-AEC. 
3) DeepVQE-S \cite{DBLP:conf/interspeech/RisteaISPGC23}, a state-of-the-art end-to-end AEC model with significantly higher computational complexity, serving as an upper-bound comparison.
Table \ref{table:exp} presents the results of the comparative experiments. It is evident that EchoFree not only outperforms ULCNet-AER and Ernst et al. proposed AEC model by a substantial margin with a lower computational load and fewer parameters, but also surpasses DeepVQE-S with significantly more parameters and MACs in terms of EchoMOS for the ST FE scenarios and DegMOS for the ST NE scenarios.

It is worth noting that to ensure a fair comparison, we retrain all reference models using our training dataset. Additionally, we maintain consistent STFT configurations across all methods, as specified in Section 3.1. For cascaded approaches integrating linear filtering with neural post filter, we utilize the same linear filtering algorithm detailed in Section 3.1 as the preprocessing step. We reproduce ULCNet-AER following the description in \cite{DBLP:conf/iwaenc/ShetuDAHM24}. For the Bark-scale feature-based AEC model, we extract features using 100 Bark-scale filters while keeping all other hyperparameters consistent with its original implementation. For DeepVQE-S, we implement the model according to \cite{DBLP:conf/interspeech/RisteaISPGC23}, including the alignment block and CCM block.

\subsection{Ablation Study}

We conduct an ablation study to analyze the impact of different training strategies on the proposed model. Table~\ref{table:exp} presents the results of our ablation experiments, comparing four configurations:
1) Only using linear AEC.
2) Training with the conventional gain loss function.  
3) Training with SSL loss.  
4) Two-stage training: first optimizing with SSL loss, followed by fine-tuning with a combination of SSL loss and the gain loss function.

The ablation test results are presented in Table 1. All configurations with NN module outperform the one with linear AEC only. and our findings reveal that models trained with SSL loss outperform those trained with the conventional gain loss function, particularly in EchoMOS for the DT scenarios, where a substantial improvement is observed. Furthermore, the two-stage training strategy yields superior results compared to direct SSL loss training. While a minor reduction in EchoMOS is observed under DT scenarios, all other evaluation metrics demonstrate notable improvements, validating the effectiveness of our proposed training strategy.

To evaluate the effectiveness of the proposed two-stage training strategy, we visualize the model outputs at each stage. As shown in \figurename~\ref{fig:visualization}, the model trained in the first stage is capable of initially suppressing the echo; however, noticeable distortions occasionally occur in the high-frequency regions. These issues are substantially mitigated after the second-stage training.

\section{Conclusion}

In this paper, we propose EchoFree, an ultra lightweight AEC model that supports streaming inference. Our model employs a cascaded framework integrating linear filtering with neural post filter. The neural post filter utilizes Bark-scale features as input and is trained using a two-stage optimization strategy based on SSL loss. Experimental results demonstrate that EchoFree achieves superior performance while maintaining a lower parameter count and reduced computational complexity compared to state-of-the-art low-complexity AEC models. In future work, we will focus on further improving EchoFree’s performance and extending its capabilities to end-to-end joint acoustic echo cancellation and noise suppression.

\bibliographystyle{IEEEtran}
\bibliography{mybib}

\end{document}